\documentclass[aps,preprint,superscriptaddress,groupedaddress,nofootinbib]{revtex4}

\usepackage{graphicx}
\usepackage[normalem]{ulem}
\usepackage{multirow}
\usepackage[export]{adjustbox}
\usepackage{float}
\usepackage{amsmath}
\usepackage{amssymb}
\usepackage[rightcaption]{sidecap}
\usepackage{hyperref}
\usepackage{xcolor}
\graphicspath{ {./Figures/} }

\def\figureautorefname~#1\null{Fig.\,#1\null}
\def\tableautorefname~#1\null{Tab.\,#1\null}

\def\equationautorefname~#1\null{Eq.\,(#1)\null}

\allowdisplaybreaks[4]

\newcommand\snowmass{\begin{center}\rule[-0.2in]{\hsize}{0.01in}\\\rule{\hsize}{0.01in}\\
    \vskip 0.1in Submitted to the  Proceedings of the US Community Study\\
    on the Future of Particle Physics (Snowmass 2021)\\
    \rule{\hsize}{0.01in}\\\rule[+0.2in]{\hsize}{0.01in} \end{center}}

\begin{document}

\title{ Tree-level Interference in VBF production of $Vh$}
\author{Chaitanya Paranjape}
\email{chaitanyaparanjape614@gmail.com}
\affiliation{Department of Physics, Indian Institute of Technology (Indian School of Mines)\\ Dhanbad, Jharkhand-826004, India}
\affiliation{Ottawa-Carleton Institute for Physics, Carleton University\\ 1125 Colonel By Drive, Ottawa, Ontario K1S 5B6, Canada}
\author{Daniel Stolarski}
\email{stolar@physics.carleton.ca}
\affiliation{Ottawa-Carleton Institute for Physics, Carleton University\\ 1125 Colonel By Drive, Ottawa, Ontario K1S 5B6, Canada}
\author{Yongcheng~Wu}
\email{ycwu0830@gmail.com}
\affiliation{Department of Physics, Oklahoma State University, Stillwater, OK, 74078, USA}

\begin{abstract}
    We study the production of the Higgs in a association with a vector ($V=W,Z$) via the VBF process, VBF-VH. In the Standard Model (SM), this process exhibits tree-level destructive interference between between $W$ and $Z$ mediated processes and is thus very sensitive to deviations in Higgs couplings to vector bosons. We study this process at both the HL-LHC as well as future high energy lepton colliders. We show in particular that the scenario where Higgs couplings have the same magnitude but opposite relative sign as in the SM, a scenario that is very difficult to distinguish without interference, can be probed with this process at either collider.\\
    \noindent {\large \bf Thematic Areas:}

\noindent $\blacksquare$ (EF01) EW Physics: Higgs Boson properties and couplings \\
\noindent $\blacksquare$ (EF04) EW Precision Physics and constraining new physics \\
\end{abstract}
\snowmass
\maketitle



\newpage

\section{Motivation}

The experimental study of the Higgs boson is well underway. As yet, its properties are consistent with the predictions of the Standard Model (SM), but more detailed study could potentially reveal new physics. As is well known, the Higgs boson tames longitudinal scattering of gauge bosons at high energy: in the absence of the Higgs, the process $VV\rightarrow VV$ grows with energy (where $V=Z,W$)~\cite{LlewellynSmith:1973yud,Veltman:1976rt,Lee:1977yc,Lee:1977eg,Passarino:1985ax,Passarino:1990hk}.
Processes involving the Higgs itself can also grow with energy if the Higgs properties differ from that of the SM. In this work we will study:
\begin{equation}
VV \rightarrow Vh, \; \; V=W,Z.
\label{eq:hardProcess}
\end{equation}
This process is sensitive to the ratio of the coupling of the Higgs to the $W$ relative to that of the $Z$. If we define $\kappa_W$ ($\kappa_Z$) as the deviation of the $W$ ($Z$) coupling to the Higgs from the SM prediction ($\kappa_W = \kappa_Z = 1$ in the SM), then we can define:
\begin{equation}
\lambda_{WZ} = \frac{\kappa_W}{\kappa_Z},
\label{eq:lambda}
\end{equation}
as the specified ratio. The process in~\autoref{eq:hardProcess} exhibits \textit{tree-level} interference effects between $W$ and $Z$ mediated processes, and the matrix element has a term that grows with energy proportional to $\lambda_{WZ} - 1$.

 One particularly interesting scenario is that when $\lambda_{WZ}$ is negative relative to the SM prediction. Tree-level processes without interference effects such as decays of $h\rightarrow ZZ^*$~\cite{Sirunyan:2017exp,Aaboud:2017oem} and $h\rightarrow WW^*$~\cite{Sirunyan:2018egh,Aaboud:2018jqu} are only sensitive to $|\lambda_{WZ}|$. Fits to the couplings by the experimental collaborations~\cite{Khachatryan:2016vau,Sirunyan:2018koj,Aad:2019mbh} measure $\lambda_{WZ}$ with approximately 10\% precision but have almost no discriminating power between positive and negative values of $\lambda_{WZ}$.\footnote{The 13 TeV CMS analysis~\cite{Sirunyan:2018koj} actually has a best fit value that is negative, and
the 13 TeV ATLAS analysis~\cite{Aad:2019mbh} does not consider negative values of $\lambda_{WZ}$. }
The ultimate LHC sensitivity on this ratio is projected to be about 2\%~\cite{Cepeda:2019klc}, but as far as we are aware, there has been no study on the sensitivity to the sign from rate measurements at the LHC. Negative values of $\lambda_{WZ}$ can arise in models with scalars that have higher isospin representations~\cite{Low:2010jp} such as the Georgi-Machacek~\cite{Georgi:1985nv} model.

Because a heavy gauge boson collider is not feasible, the gauge boson scattering is studied experimentally via vector boson fusion (VBF), where $W$ or $Z$'s are radiated off the initial state fermions and then scatter off one another. In this report we study VBF production of $Vh$ at both the HL-LHC and a future high energy lepton collider.

\section{Study at the LHC}


At the LHC, the VBF process means producing $Vh$ in association with two jets with a large rapidity gap. Here we focus on the process
\begin{equation}
    p p \rightarrow Z h j j
    \label{lhcProcess}
\end{equation}
where $j$ is a jet. We leave the process with a $W$ in the final state to future work. We also use the leptonic decay of the $Z$ and the $b\bar{b}$ decay of the Higgs:
\begin{eqnarray}
    Z &\rightarrow& \ell^+ \ell^- \nonumber\\
    h &\rightarrow& b\bar{b}
\end{eqnarray}
The signal cross section is already quite small (see \autoref{eq:signalCSX} below), therefore the Higgs mode with the largest branching ratio is chosen,
while the $Z$ mode allows for significant background reduction. The decay of the $Z$ to neutrinos could also be a viable signal because it has a larger rate than charged leptons, but it is not considered here.

The signal is simulated using {\tt MadGraph5\_aMC@NLO}~\cite{Alwall:2014hca}, and the $QCD = 0$ flag is used to ensure VBF topology. The same process without the $QCD=0$ flag is a (subdominant) background. The events are hadronized using {\tt PYTHIA8}~\cite{Sjostrand:2007gs} and {\tt Delphes}~\cite{deFavereau:2013fsa} is used as a detector simulation.
The tree-level signal cross section (before taking branching ratios into account) scales with the coupling modifiers as
\begin{equation}
\sigma_{\text{sig}} \sim (17.41\  \mbox{fb}) \cdot \kappa^2_W - (14.755\  \mbox{fb}) \cdot \kappa_W \kappa_Z + (12.41\  \mbox{fb}) \cdot \kappa^2_Z  \label{eq:signalCSX}
\end{equation}
From this we see that there is large destructive interference in the SM ($\kappa_W = \kappa_Z = 1$), and that if $\kappa_W = - \kappa_Z=\pm 1$, then the cross section is enhanced by a factor $\sim 3$.

The dominant backgrounds for this analysis are:
\begin{eqnarray}
    p p &\rightarrow& Z b \bar{b} j j \nonumber\\
    p p &\rightarrow& t \bar{t}     \label{eq:bcgLHC}
\end{eqnarray}
where in the first process the $Z$ decays leptonically and in the second both tops decay leptonically. The background cross section for the first (second) process is roughly 1 (5) pb with branching ratios taken into account, while the signal cross section is about 1 fb. Therefore, cuts that significantly enhance the signal to background ratio are needed. Backgrounds with two $Z$ bosons are also considered but are subdominant.

Our analysis is built from characteristic VBF cuts by identifying the forward-backward jet pair with highest invariant mass as the VBF-tagging jets. The final set of cuts we developed are:

$\bullet$ At least one forward-backward jet pair must exist.

$\bullet$ All jets under consideration must have $p_T \geq 20\ \mbox{GeV},\ |\eta|\leq 5$.

$\bullet$ Number of $B$-tagged Jets $\geq \ 2$.

$\bullet$ Number of jets that are VBF AND $B$-tagged $= \ 0$.

$\bullet$ Invariant mass of the detected OSSF Lepton pair $\in \ (81 \,\mbox{GeV},101\, \mbox{GeV})$.

$\bullet$ $|\eta_1 - \eta_2|$ of VBF-tagged Jets $\geq 4$.

$\bullet$ Total invariant mass of VBF-Tagged jet pair $\geq 1000\ \mbox{GeV}$.

$\bullet$ Missing $E_T$ $< 50\ \mbox{GeV}$.

$\bullet$ Jet-1 $p_T$ $\geq 100\ \mbox{GeV}$.

$\bullet$ Jet-2 $p_T$ $\geq 70\ \mbox{GeV}$.

$\bullet$ Jet-3 $p_T$ $\geq 50\ \mbox{GeV}$.

$\bullet$ $B$-Jet-1 $p_T$ $\geq 55\ \mbox{GeV}$.

$\bullet$ $B$-Jet-2 $p_T$ $\geq 55\ \mbox{GeV}$.

$\bullet$ $\Delta R_{b \bar{b}} = \sqrt{(\Delta \eta)^2 + (\Delta \phi)^2} $ $\leq 2$.

$\bullet$ Higgs mass reconstructed with standard jets $\in (110\ \mbox{GeV},130\ \mbox{GeV})$.

$\bullet$ Higgs mass reconstructed with BDRS algorithm~\cite{Butterworth:2008iy} with $R=2.0$ $\in (110\ \mbox{GeV},130\ \mbox{GeV})$. \\
The yields and related data are displayed in~\autoref{tab:1}. We see that the signal to background ratio has increased significantly, but that detection with the HL-LHC at 3,000 fb$^{-1}$ will still be challenging.

\begin{figure}[h!]
  \centering
  \includegraphics[width=0.6\textwidth]{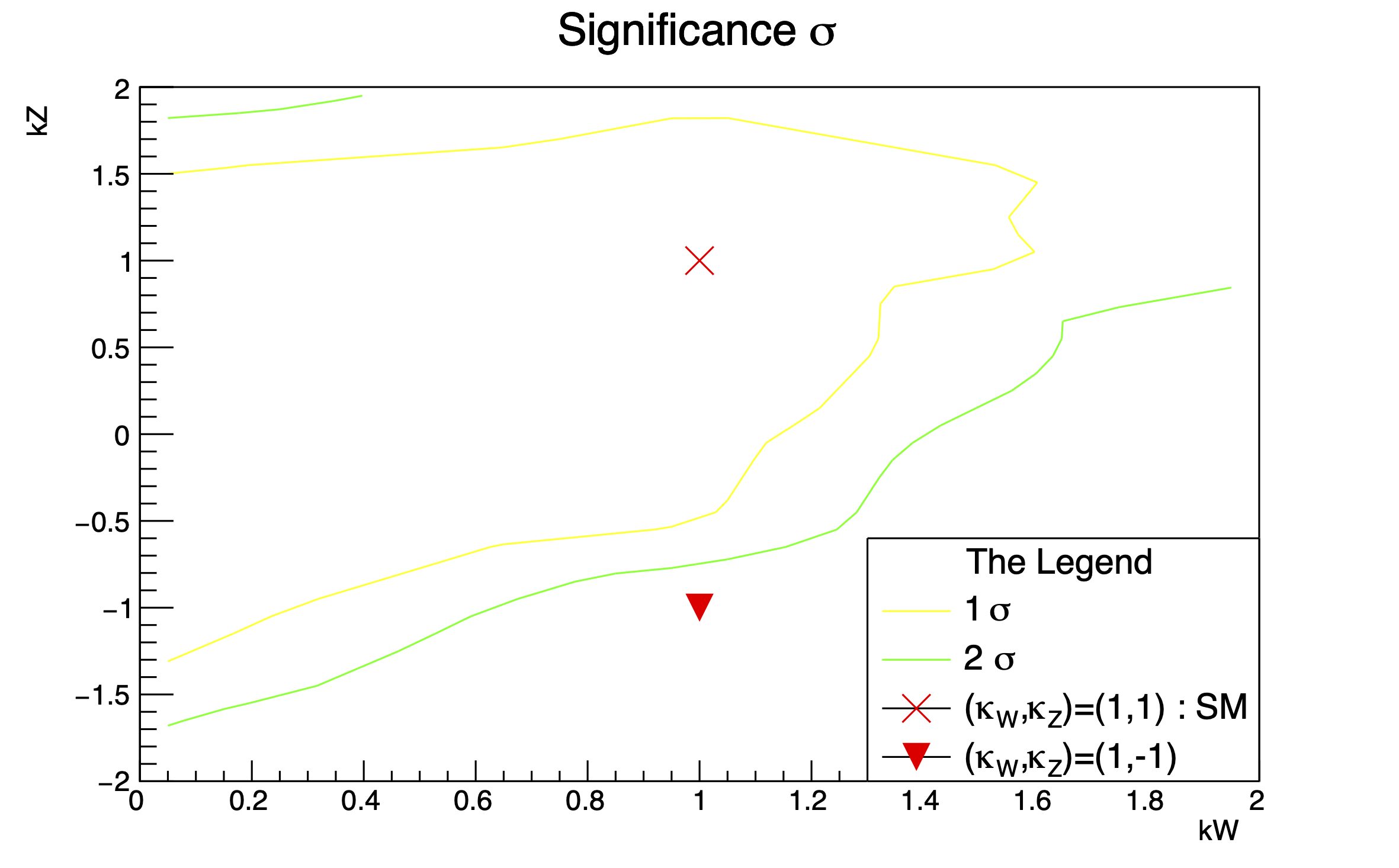}
  \caption{1- and 2-$\sigma$ sensitivity of this measurement to $\kappa_z$ and $\kappa_W$. The x is the SM point, while the the triangle is the one with the same magnitude but opposite relative sign which is weakly constrained with current data.  }
  \label{fig:KF5}
\end{figure}

All of the steps so far had been performed using the SM with $\kappa_W=\kappa_Z=1$. We can now vary the values of $(\kappa_W,\kappa_Z)$ and determine the sensitivity to those parameters. The kinematic distributions are also changed and the change of efficiency for non-SM points is taken into account. We estimate the significance as follows as a function of $\kappa_{W/Z}$ assuming an observation at the SM expectation by:
\begin{equation}
     \sigma = \frac{|A(\kappa_W,\kappa_Z) - A_{SM}|}{\sqrt{A_{SM} + (\beta \cdot A_{SM})^2}}
     \label{eq:significance}
\end{equation}
where $A=S+B$ is the total signal+background yield at the corresponding point. The errors in the denominator are statistical and systematic respectively, and we have parameterized our systematic errors with the parameter $\beta$ which we take to be $\beta=0.1$.
Our results are given in~\autoref{fig:KF5}, where the SM point is shown with a cross and the point with $(\kappa_W,\kappa_Z)=(1,-1)$ is shown with a triangle and is expected to be excluded at more than 2$\sigma$.

\begin{table}
    \centering
    \begin{tabular}{ |p{2.5cm}|p{5cm}|p{4cm}|  }
 \hline
 Process & MC selection efficiency & \# events (3,000 fb$^{-1}$) \\
 \hline
 Signal & $503\ \ /100$k         & $6.23 \ \pm\  0.28$ \\
 \hline
 $Zb\bar{b}jj$ & $9\ \ \ \ /700$k & $42.93 \ \pm\  14.31$ \\
 \hline
  $t\bar{t}$  & $3\ \ \ \ \ /5$M & $9.56 \ \pm\  5.52$ \\
 \hline
\end{tabular}
    \caption{Monte Carlo selection efficiency with the number of events generated in the denominator in the second column. The third column is the number of expected events at the HL-LHC, and uncertainties are due to Monte Carlo statistics.}
    \label{tab:1}
\end{table}

\section{Study at a future high energy lepton collider}

The same type of measurement can be made at a lepton collider, where the processes are:
\begin{align}
e^+\ e^- &\to \nu_e\ \bar{\nu}_e\ Z\ h, \nonumber\\
e^+\ e^- &\to \nu_e\ e\ W\ h.
\label{eq:VBFprocess}
\end{align}
At a lepton collider, the signal to background ratio is much more favourable than at the LHC.
This measurement was studied in~\cite{Stolarski:2020qim} and we here give a brief summary of the results. As with~\autoref{lhcProcess}, the processes in~\autoref{eq:VBFprocess} grows with the center of mass energy if there are deviations of the Higgs couplings from their SM predictions. Therefore a high energy lepton collider will have the best sensitivity. Here we analyze the CLIC $e^+e^-$ machine as a benchmark, but the same analysis would apply to a high energy muon collider.

These two processes also possess significant destructive interference between $W$ and $Z$ mediated contributions. For several benchmark collider scenarios, we show in~\autoref{tab:CSIndividual} the cross sections for $Wh$ and $Zh$ processes with
\begin{equation}
\sigma_{\rm total} = \kappa_W^2\sigma_W + \kappa_W\kappa_Z\sigma_{WZ}+\kappa_Z^2\sigma_Z.
\label{eq:sigma}
\end{equation}
From~\autoref{tab:CSIndividual}, one can see that the interference effect is even larger than the individual contribution of $\sigma_W$ and $\sigma_Z$. Utilizing such effect can have a quite sensitive measurement of $\kappa_W$, $\kappa_Z$ and $\lambda_{WZ}$.

\begin{table}[!hbt]
    \centering
    \begin{tabular}{c|c|rr|rr}
    \hline\hline
    \multicolumn{2}{c|}{$\sigma$ [fb]} & \multicolumn{2}{c|}{$Wh$} & \multicolumn{2}{c}{$Zh$} \\
    \cline{1-2}
    $\sqrt{s}$ [GeV] &  & $P(e^-)=-80\%$ & $P(e^-)=80\%$ & $P(e^-)=-80\%$ & $P(e^-)=80\%$\\
    \hline
    \multirow{3}{*}{1500} & $\sigma_Z$ & $8.25\times10^{0}$ & $3.18\times10^{0}$ & $3.85\times10^{0}$ & $4.25\times10^{-1}$ \\
    & $\sigma_W$ & $1.22\times10^{1}$ & $4.11\times10^{0}$ & $6.85\times10^{0}$ & $7.66\times10^{-1}$ \\
    & $\sigma_{WZ}$ & $-1.28\times10^{1}$ & $-5.46\times10^{0}$ & $-5.38\times10^{0}$ & $5.93\times10^{-1}$ \\
    \hline
    \multirow{3}{*}{3000} & $\sigma_Z$ & $3.51\times10^{1}$ & $1.34\times10^{1}$ & $1.87\times10^{1}$ & $2.09\times10^{0}$ \\
    & $\sigma_W$ & $4.31\times10^{1}$ & $1.50\times10^{1}$ & $2.97\times10^{1}$ & $3.27\times10^{0}$ \\
    & $\sigma_{WZ}$ & $-6.32\times10^{1}$ & $-2.52\times10^{1}$ & $-3.13\times10^{1}$ & $-3.45\times10^{0}$ \\
    \hline
    \hline
    \end{tabular}
    \caption{The individual contributions to total cross section in the parameterization of~\autoref{eq:sigma} for $Wh$ and $Zh$ VBF processes at different collision energies and different polarizations. The cross section is obtained from {\tt MadGraph5\_aMC@NLO} with cuts: $p_T^\ell>10$ GeV and $|\eta_\ell|<3.5$. The polarization configuration is following~\cite{Roloff:2018dqu}.}
    \label{tab:CSIndividual}
\end{table}

In the phenomenological studies, we consider the leptonic decay of gauge bosons ($Z\to\ell\ell, W\to\ell\nu$), and $h\to b\bar b$. The dominant backgrounds for this analysis are
\begin{subequations}
    \begin{align}
    &e^- e^+ \to e^{\pm} \nu_{e} W^\mp Z \to e^\pm \nu_e \ell^\mp \nu_\ell b\bar{b},\\
    &e^- e^+ \to \nu_e \bar{\nu}_e Z Z \to \nu_e \bar{\nu}_e \ell^-\ell^+ b\bar{b}. 
    \end{align}
\end{subequations}
Other backgrounds were also considered, but they are comparitavely unimportant.
The events are generated using {\tt MadGraph5\_aMC@NLO}~\cite{Alwall:2014hca} with {\tt PYTHIA8}~\cite{Sjostrand:2007gs}  used for showering and hadronization. The detector effects are simulated with {\tt Delphes}~\cite{deFavereau:2013fsa} using the CLIC card~\cite{Leogrande:2019qbe}. In order to improve the sensitivity, we simulate both 3 TeV and 1.5 TeV events with $P(e^-)=-0.8$ for the electron beam which are two scenarios for CLIC with 4000 and 2000 fb$^{-1}$ luminosity respectively~\cite{Roloff:2018dqu}.

\begin{table}[!tbp]
    \centering
    \begin{tabular}{|c|c|c|}
    \hline
    Cuts & $Wh$-Cuts & $Zh$-Cuts \\ \hline
    \multirow{3}{*}{Basic Cuts} & \multicolumn{2}{c|}{$p_T^\ell > 20$ GeV, $N_\ell = 2$} \\
        & \multicolumn{2}{c|}{$p_T^j>20$ GeV, $N_b=2$} \\ \cline{2-3}
        & $N_e\geq1$ & 1 OSSF Pair \\ \hline
    $m_{bb}$ & \multicolumn{2}{c|}{$95\ {\rm GeV}\leq m_{bb}\leq130\ {\rm GeV}$} \\ \hline
    $m_{\ell\ell}$ & $m_{\ell\ell}\leq80\ {\rm GeV}$ or $m_{\ell\ell}\geq98\ {\rm GeV}$ & $75\ {\rm GeV}\leq m_{\ell\ell}\leq 100\ {\rm GeV}$ \\ \hline
    $H_T$ & $\begin{cases}H_T\leq 2500\ {\rm GeV} & \sqrt{s} = 3000\ {\rm GeV}\\ H_T\leq 1100\ {\rm GeV}& \sqrt{s} = 1500\ {\rm GeV}\end{cases}$ & $\begin{cases}H_T\leq 1500\ {\rm GeV} & \sqrt{s} = 3000\ {\rm GeV}\\ H_T\leq 700\ {\rm GeV}& \sqrt{s} = 1500\ {\rm GeV}\end{cases}$ \\ \hline
    \end{tabular}%
    \caption{The Cuts used for $Wh$ channel and $Zh$ channel.}
    \label{tab:cuts_wh_zh}
\end{table}

\begin{table}[!tbp]
    \centering
    \resizebox{\textwidth}{!}{%
    \begin{tabular}{cccccccc}
    \hline\hline
    \multicolumn{2}{c}{\multirow{2}{*}{$\sigma$ (fb)}} & \multicolumn{3}{c}{$\sqrt{s} = 3.0$ TeV, $\mathcal{L}=4$ ab$^{-1}$} & \multicolumn{3}{c}{$\sqrt{s} = 1.5$ TeV $\mathcal{L}=2$ ab$^{-1}$} \\
    \multicolumn{2}{c}{} & Before Cuts & $Wh$-Cuts & $Zh$-Cuts & Before Cuts & $Wh$-Cuts & $Zh$-Cuts \\ \hline\hline
    \multirow{2}{*}{Signal} & $Wh$(VBF) & $1.97\times10^0$ & $7.26\times10^{-2}$ & $1.36\times10^{-3}$ & $9.62\times10^{-1}$ & $6.54\times10^{-2}$ & $2.37\times10^{-3}$ \\
        & $Zh$(VBF) & $6.47\times10^{-1}$ & $3.49\times10^{-3}$ & $7.21\times10^{-2}$ & $2.03\times10^{-1}$ & $1.30\times10^{-3}$ & $2.87\times10^{-2}$ \\ \hline
    \multirow{2}{*}{BG} 
        & $WZ$(VBF) & $4.47\times10^0$ & $9.97\times10^{-3}$ & $2.16\times10^{-4}$ & $1.84\times10^0$ & $5.86\times10^{-3}$ & $1.96\times10^{-4}$ \\
        & $ZZ$(VBF) & $1.92\times10^0$ & $4.21\times10^{-4}$ & $8.07\times10^{-3}$ & $5.92\times10^{-1}$ & $1.48\times10^{-4}$ & $2.88\times10^{-3}$ \\
    \hline
        & & Precision (\%) & 6.18 &  6.17 & Precision (\%) &  9.53 &   13.5 \\
        \hline\hline
    \end{tabular}%
    }
    \caption{The cross sections for signal and the dominant background (BG) processes (with final states $b\bar{b}\ell^+\ell^-$ or $b\bar{b}\ell\nu$) at $\sqrt{s} = 1500,\ 3000$ GeV for $P(e^-) = -0.8$. Note that for the VBF processes, $p_T^\ell > 10$ GeV and $|\eta^\ell|<3.5$ are imposed at the generation level for the forward/backward charged lepton. We also quote the the precision on the measurement of signal cross section that can be extracted with the given luminosity.}
    \label{tab:allcs}
\end{table}

To suppress the background, we implement several selection cuts to single out the signal events which are listed in~\autoref{tab:cuts_wh_zh}. The cut flow for signal (assuming $\kappa_W=1$ and $\kappa_Z=1$) and background processes are further listed in~\autoref{tab:allcs}. By assuming that the selection efficiency will not change significantly for different values of $\kappa_{W,Z}$, we can directly obtain the signal events for all other cases by proper scaling according to $\kappa_{W,Z}$. We can then compute the log-liklihood function of $\kappa_W$ and $\kappa_Z$ to get confidence contours of those parameters.

\begin{figure}
    \centering
    \includegraphics[width=0.31\textwidth]{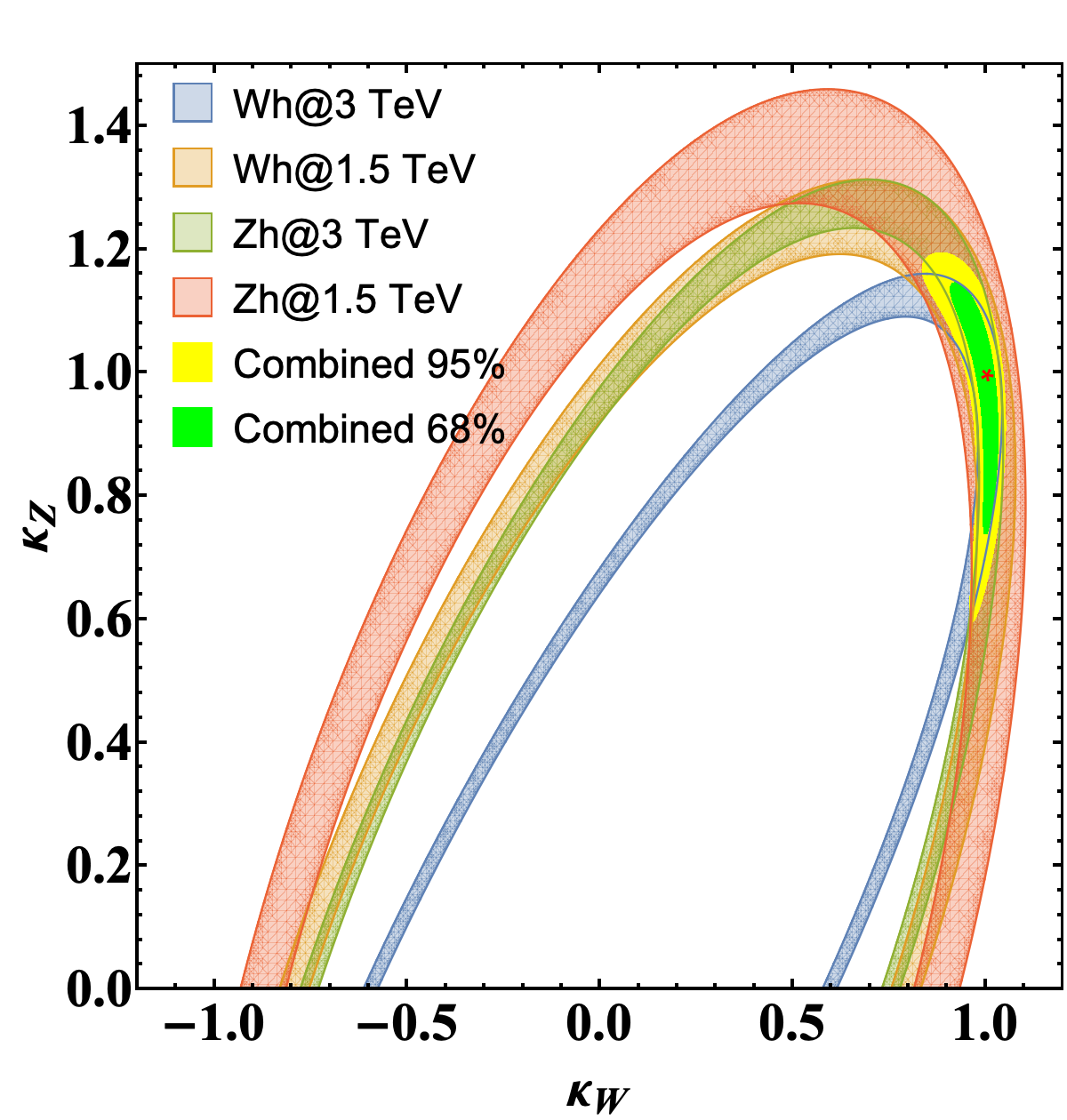}
    \includegraphics[width=0.32\textwidth]{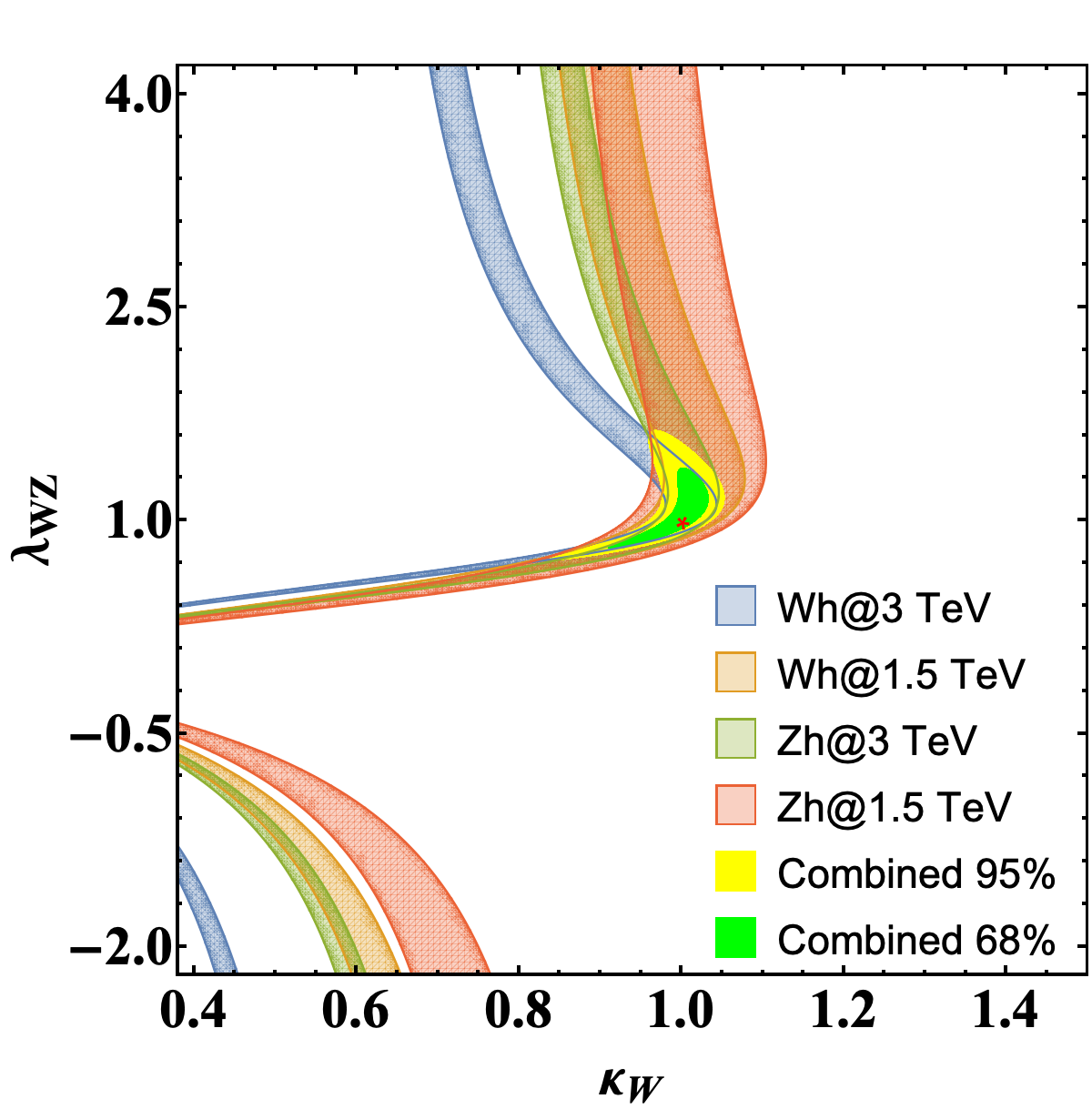}
    \includegraphics[width=0.32\textwidth]{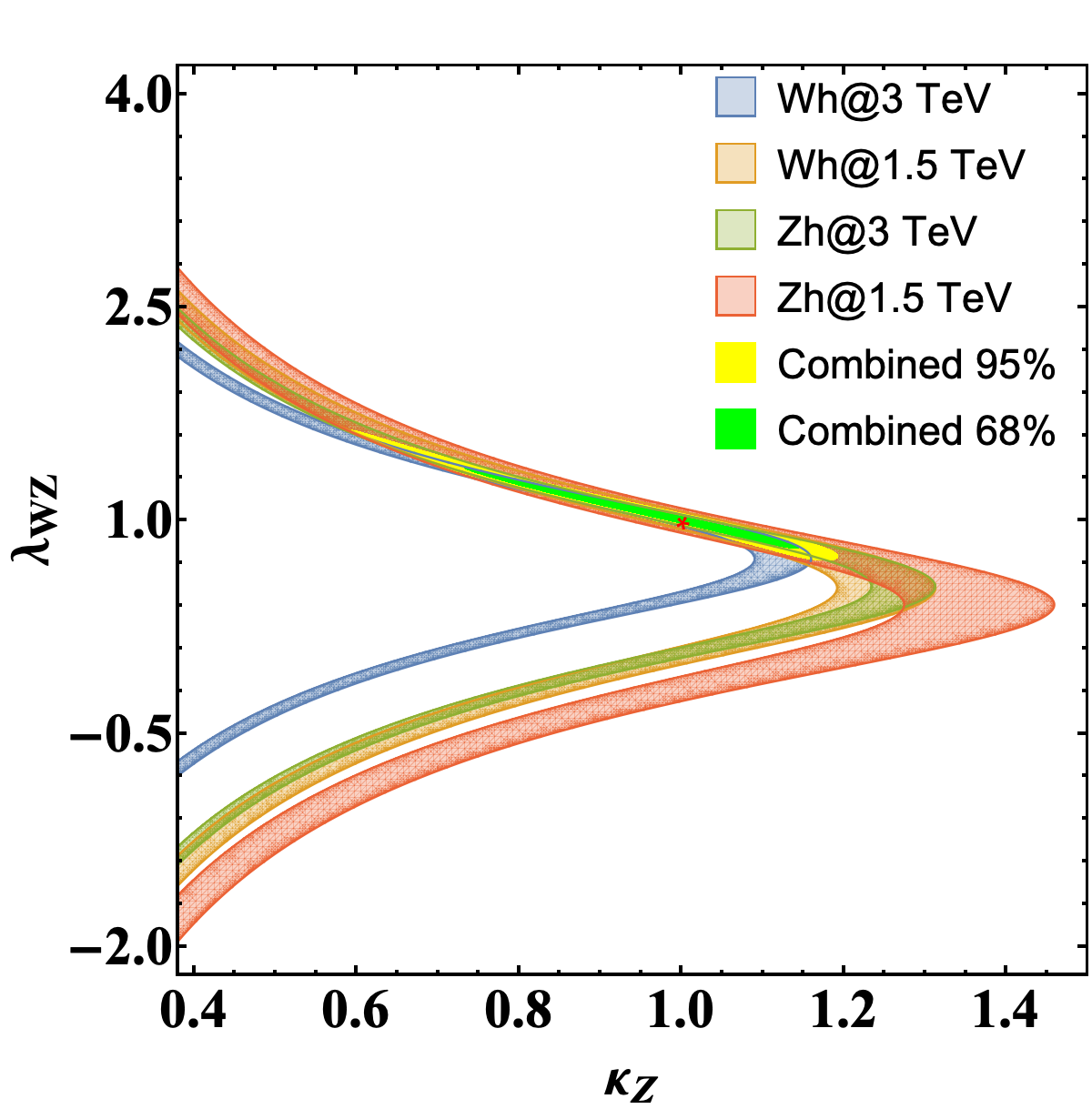}
    \caption{The constraints in the $\kappa_W$-$\kappa_Z$, $\kappa_W$-$\lambda_{WZ}$, and $\kappa_Z$-$\lambda_{WZ}$ planes from the total rate measurements. We show the contours from the four different measurements at 68\%, and also show the combined constraints at 68\% C.L. (95\% C.L.) in green (yellow). The SM values are indicated as red points.
    }
    \label{fig:LamWZtotalrate}
\end{figure}

Combining all the channels, we can get the 68\% and 95\% confidence level regions which are shown in~\autoref{fig:LamWZtotalrate}. We see that the combination can restrict the allowed region to be much smaller than for any individual measurement.
Further, we can also estimate the luminosity that is needed to exclude some non-SM benchmark points at specific scenarios. The results are presented in~\autoref{tab:lumExclude} which shows that with significantly less data, we can exclude those BSM scenarios. Besides the total rate, we can further utilize the distribution to obtain stronger sensitivities.

\begin{table}
    \centering
    \begin{tabular}{ccc}
        \hline\hline
        Benchmark & $\sqrt{s} = 3.0$ TeV & $\sqrt{s} = 1.5$ TeV\\\hline\hline
        $\kappa_W = \pm1$, $\kappa_Z = \mp1$ & 3.4 fb$^{-1}$ & 14.1 fb$^{-1}$ \\
        $\kappa_W = 1$, $\kappa_Z = 0 $ & 29.3 fb$^{-1}$ & 243.3 fb$^{-1}$ \\
        $\kappa_W = 0$, $\kappa_Z = 1 $ & 62.1 fb$^{-1}$ & 1772.4 fb$^{-1}$ \\\hline\hline
    \end{tabular}
    \caption{The luminosity that is needed to exclude specific benchmark points at 95\% C.L. against the SM case ($\kappa_W=1$ and $\kappa_Z=1$).}
    \label{tab:lumExclude}
\end{table}

\bibliographystyle{JHEP}
\bibliography{references}

\providecommand{\href}[2]{#2}\begingroup\raggedright\begin{thebibliography}{10}

\bibitem{LlewellynSmith:1973yud}
C.~H. Llewellyn~Smith, {\it {High-Energy Behavior and Gauge Symmetry}},  {\em
  Phys. Lett.} {\bf 46B} (1973) 233--236.

\bibitem{Veltman:1976rt}
M.~J.~G. Veltman, {\it {Second Threshold in Weak Interactions}},  {\em Acta
  Phys. Polon.} {\bf B8} (1977) 475.

\bibitem{Lee:1977yc}
B.~W. Lee, C.~Quigg, and H.~B. Thacker, {\it {The Strength of Weak Interactions
  at Very High-Energies and the Higgs Boson Mass}},  {\em Phys. Rev. Lett.}
  {\bf 38} (1977) 883--885.

\bibitem{Lee:1977eg}
B.~W. Lee, C.~Quigg, and H.~B. Thacker, {\it {Weak Interactions at Very
  High-Energies: The Role of the Higgs Boson Mass}},  {\em Phys. Rev.} {\bf
  D16} (1977) 1519.

\bibitem{Passarino:1985ax}
G.~Passarino, {\it {Large Masses, Unitarity and One Loop Corrections}},  {\em
  Phys. Lett.} {\bf 156B} (1985) 231--235.

\bibitem{Passarino:1990hk}
G.~Passarino, {\it {W W scattering and perturbative unitarity}},  {\em Nucl.
  Phys.} {\bf B343} (1990) 31--59.

\bibitem{Sirunyan:2017exp}
{\bf CMS} Collaboration, A.~M. Sirunyan et~al., {\it {Measurements of
  properties of the Higgs boson decaying into the four-lepton final state in pp
  collisions at $ \sqrt{s}=13 $ TeV}},  {\em JHEP} {\bf 11} (2017) 047,
  [\href{http://arxiv.org/abs/1706.09936}{{\tt arXiv:1706.09936}}].

\bibitem{Aaboud:2017oem}
{\bf ATLAS} Collaboration, M.~Aaboud et~al., {\it {Measurement of inclusive and
  differential cross sections in the $H \rightarrow ZZ^* \rightarrow 4\ell$
  decay channel in $pp$ collisions at $\sqrt{s}=13$ TeV with the ATLAS
  detector}},  {\em JHEP} {\bf 10} (2017) 132,
  [\href{http://arxiv.org/abs/1708.02810}{{\tt arXiv:1708.02810}}].

\bibitem{Sirunyan:2018egh}
{\bf CMS} Collaboration, A.~M. Sirunyan et~al., {\it {Measurements of
  properties of the Higgs boson decaying to a W boson pair in pp collisions at
  $\sqrt{s}=$ 13 TeV}},  {\em Phys. Lett.} {\bf B791} (2019) 96,
  [\href{http://arxiv.org/abs/1806.05246}{{\tt arXiv:1806.05246}}].

\bibitem{Aaboud:2018jqu}
{\bf ATLAS} Collaboration, M.~Aaboud et~al., {\it {Measurements of gluon-gluon
  fusion and vector-boson fusion Higgs boson production cross-sections in the
  $H \to WW^{\ast} \to e\nu\mu\nu$ decay channel in $pp$ collisions at
  $\sqrt{s}=13$ TeV with the ATLAS detector}},  {\em Phys. Lett.} {\bf B789}
  (2019) 508--529, [\href{http://arxiv.org/abs/1808.09054}{{\tt
  arXiv:1808.09054}}].

\bibitem{Khachatryan:2016vau}
{\bf ATLAS, CMS} Collaboration, G.~Aad et~al., {\it {Measurements of the Higgs
  boson production and decay rates and constraints on its couplings from a
  combined ATLAS and CMS analysis of the LHC pp collision data at $ \sqrt{s}=7
  $ and 8 TeV}},  {\em JHEP} {\bf 08} (2016) 045,
  [\href{http://arxiv.org/abs/1606.02266}{{\tt arXiv:1606.02266}}].

\bibitem{Sirunyan:2018koj}
{\bf CMS} Collaboration, A.~M. Sirunyan et~al., {\it {Combined measurements of
  Higgs boson couplings in proton--proton collisions at $\sqrt{s}=13\,\text
  {Te}\text {V} $}},  {\em Eur. Phys. J. C} {\bf 79} (2019), no.~5 421,
  [\href{http://arxiv.org/abs/1809.10733}{{\tt arXiv:1809.10733}}].

\bibitem{Aad:2019mbh}
{\bf ATLAS} Collaboration, G.~Aad et~al., {\it {Combined measurements of Higgs
  boson production and decay using up to $80$ fb$^{-1}$ of proton-proton
  collision data at $\sqrt{s}=$ 13 TeV collected with the ATLAS experiment}},
  {\em Phys. Rev. D} {\bf 101} (2020), no.~1 012002,
  [\href{http://arxiv.org/abs/1909.02845}{{\tt arXiv:1909.02845}}].

\bibitem{Cepeda:2019klc}
M.~Cepeda et~al., {\em {Report from Working Group 2}: {Higgs Physics at the
  HL-LHC and HE-LHC}}, vol.~7, pp.~221--584.
\newblock 12, 2019.
\newblock \href{http://arxiv.org/abs/1902.00134}{{\tt arXiv:1902.00134}}.

\bibitem{Low:2010jp}
I.~Low and J.~Lykken, {\it {Revealing the Electroweak Properties of a New
  Scalar Resonance}},  {\em JHEP} {\bf 10} (2010) 053,
  [\href{http://arxiv.org/abs/1005.0872}{{\tt arXiv:1005.0872}}].

\bibitem{Georgi:1985nv}
H.~Georgi and M.~Machacek, {\it {DOUBLY CHARGED HIGGS BOSONS}},  {\em Nucl.
  Phys.} {\bf B262} (1985) 463--477.

\bibitem{Alwall:2014hca}
J.~Alwall, R.~Frederix, S.~Frixione, V.~Hirschi, F.~Maltoni, O.~Mattelaer,
  H.~S. Shao, T.~Stelzer, P.~Torrielli, and M.~Zaro, {\it {The automated
  computation of tree-level and next-to-leading order differential cross
  sections, and their matching to parton shower simulations}},  {\em JHEP} {\bf
  07} (2014) 079, [\href{http://arxiv.org/abs/1405.0301}{{\tt
  arXiv:1405.0301}}].

\bibitem{Sjostrand:2007gs}
T.~Sjostrand, S.~Mrenna, and P.~Z. Skands, {\it {A Brief Introduction to PYTHIA
  8.1}},  {\em Comput. Phys. Commun.} {\bf 178} (2008) 852--867,
  [\href{http://arxiv.org/abs/0710.3820}{{\tt arXiv:0710.3820}}].

\bibitem{deFavereau:2013fsa}
{\bf DELPHES 3} Collaboration, J.~de~Favereau, C.~Delaere, P.~Demin,
  A.~Giammanco, V.~Lemaître, A.~Mertens, and M.~Selvaggi, {\it {DELPHES 3, A
  modular framework for fast simulation of a generic collider experiment}},
  {\em JHEP} {\bf 02} (2014) 057, [\href{http://arxiv.org/abs/1307.6346}{{\tt
  arXiv:1307.6346}}].

\bibitem{Butterworth:2008iy}
J.~M. Butterworth, A.~R. Davison, M.~Rubin, and G.~P. Salam, {\it {Jet
  substructure as a new Higgs search channel at the LHC}},  {\em Phys. Rev.
  Lett.} {\bf 100} (2008) 242001, [\href{http://arxiv.org/abs/0802.2470}{{\tt
  arXiv:0802.2470}}].

\bibitem{Stolarski:2020qim}
D.~Stolarski and Y.~Wu, {\it {Tree-level interference in vector boson fusion
  production of Vh}},  {\em Phys. Rev. D} {\bf 102} (2020), no.~3 033006,
  [\href{http://arxiv.org/abs/2006.09374}{{\tt arXiv:2006.09374}}].

\bibitem{Roloff:2018dqu}
{\bf CLIC, CLICdp} Collaboration, P.~Roloff, R.~Franceschini, U.~Schnoor, and
  A.~Wulzer, {\it {The Compact Linear e$^+$e$^-$ Collider (CLIC): Physics
  Potential}},  \href{http://arxiv.org/abs/1812.07986}{{\tt arXiv:1812.07986}}.

\bibitem{Leogrande:2019qbe}
E.~Leogrande, P.~Roloff, U.~Schnoor, and M.~Weber, {\it {A DELPHES card for the
  CLIC detector}},  \href{http://arxiv.org/abs/1909.12728}{{\tt
  arXiv:1909.12728}}.

\end{thebibliography}\endgroup

\end{document}